\documentclass[aps,prl,twocolumn,floatfix,showpacs,superscriptaddress]{revtex4-1}
\usepackage{amsmath,amsfonts,amssymb,bm}
\usepackage[utf8]{inputenc}
\usepackage{graphicx}
\usepackage{color}
\usepackage{subfigure}
\usepackage{multirow}%for tabular
\usepackage{textcomp}%for +-
\usepackage{slashed}

\definecolor{purple}{rgb}{0.5,0,0.5}
\definecolor{blue}{rgb}{0.0,0,1.0}

\usepackage[colorlinks=true, pdfstartview=FitV, linkcolor=purple, citecolor= purple, urlcolor=blue]{hyperref}
\begin{document}

%\preprint{}

\title{Pion Parton Distribution Function in Light-Front Holographic QCD}

\author{Lei Chang}
\affiliation{School of Physics, Nankai University, Tianjin 300071, China}
%\email{leichang@nankai.edu.cn}

\author{Kh\'epani Raya}
\affiliation{School of Physics, Nankai University, Tianjin 300071, China}
%\email{khepani@nankai.edu.cn}

\author{Xiaobin Wang}
\affiliation{School of Physics, Nankai University, Tianjin 300071, China}

\date{\today}

\begin{abstract}
The valence-quark distribution function of the pion has been of interest for decades; particularly, the profile it should adopt when $x\to1$ (the large-$x$ behavior) is subject of a long-standing debate. In the light-front holographic QCD (LFHQCD) approach, this behavior is controlled by the so called reparametrization function, $w_\tau(x)$, which is not fully determined from first principles. We show that, owing to the flexibility of $w_\tau(x)$, the large-$x$ profile $u^{\pi}(x)\sim (1-x)^{2}$  can be contained within the LFHQCD formalism. This is in contrast with a previous LFHQCD study~(Guy F. de Teramond et al., Phys. Rev. Lett., 120(18), 2018) in which $u^{\pi}(x)\sim (1-x)^{1}$ is found instead. Given our observations, augmented by perturbative QCD and recent lattice QCD results, we state that the large-$x$ exponent of ``2'' cannot be excluded.
\end{abstract}

%
%\pacs{12.38.Aw, 11.10.St, 13.40.Hq, %decay electromagnetic (particle physics), 13.40.Hq 14.40.Lb %charmed mesons, 14.40.Lb }
%\keywords{}

\maketitle

  \noindent\emph{Motivation}\,---\,
  During the development of parton models, around the 1970s, a connection between the proton electromagnetic form factors (obtained via exclusive process) and its structure functions (inferred from deep inelastic scattering) was realized by Drell-Yann~\cite{Drell:1969km} and West~\cite{West:1970av}. Their findings yielded the so-called Drell-Yan-West relation (DYW), which entails that, when the momentum transfer ($-t=Q^2$) becomes asymptotically large, the proton electromagnetic form factor (EFF) decays as
  \begin{equation}
  \label{eq:EFFDYW}
  F_{1 }^{p}(t)\sim \frac{1}{(-t)^{\tau - 1}}\;,
  \end{equation}
  while the corresponding parton distribution function (PDF) exhibits the large-$x$ (\emph{i.e.}, $x\to1$) behavior of
  \begin{equation}
  \label{eq:PDFDYW}
  u^{p}(x)\sim (1-x)^{2\tau - 3}\;.
  \end{equation}
Here, $x$ is the longitudinal momentum fraction carried by the parton (or Bjorken-$x$)~\cite{Bjorken:1968dy}, and $\tau$, called \emph{twist}, denotes the number of $\tau$-components of the hadron state. In a subsequent work by Ezawa~\cite{Ezawa:1974wm}, it was shown that the pion violates the DYW relation. This can be attributed to the different number of constituents and spin.  It is seen that, while the EFF exhibits the same asymptotic profile for both hadrons, Eq.~\eqref{eq:EFFDYW}, the pion parton distribution function adopts the large-$x$ form
\begin{equation}
\label{eq:pionPDF1}
u^\pi(x)\sim (1-x)^{2\tau - 2}\;.
\end{equation}
The leading-twist ($\tau=3$ for proton, $\tau=2$ for pion) entails the well-known $1/(-t)^2$ and $1/(-t)$ power-law-like decays of the proton and pion EFFs~\cite{Lepage:1980fj}, respectively, and the predicted $x\to1$ behavior of the PDFs is
  \begin{eqnarray}
  \label{eq:largexgood0}
  u^p(x)&\sim&(1-x)^3\;,\\
    \label{eq:largexgood}
  u^\pi(x)&\sim&(1-x)^2\;.
  \end{eqnarray}
  Those patterns are further supported by perturbative Quantum Chromodynamics (pQCD)~\cite{Farrar:1975yb,Berger:1979du,Lepage:1980fj}. It is worth clarifying that Eqs.~\eqref{eq:largexgood0}-\eqref{eq:largexgood} are valid at a certain energy scale that marks the boundary between strong and perturbative dynamics~\cite{Cui:2020dlm,Ding:2019lwe,Ding:2019qlr}. Above this scale, the anomalous dimensions increase logarithmically and so the large-$x$ exponents~\cite{Altarelli:1977zs}. 
  
  Assuming a theory in which the quarks interact via the exchange of a vector-boson, asymptotically damped as $(1/k^2)^\beta$,  the pion case in Eq.~\eqref{eq:largexgood0} generalizes as~\cite{Holt:2010vj}:
  \begin{equation}
u^\pi(x)\sim (1-x)^{2\beta}\;.
  \end{equation}
  Hence, the large-$x$ behavior of the valence-quark PDF is a direct measure of the momentum-dependence of the underlying interaction~\cite{Farrar:1975yb,Berger:1979du,Holt:2010vj,Hecht:2000xa}. 
  
  In the novel approach of light-front holographic QCD (LFHQCD)~\cite{Brodsky:2014yha,Zou:2018eam}, it is suggested that the DYW relation is preserved for both the proton and pion~\cite{deTeramond:2018ecg}. Thereby, this framework predicts a valence pion PDF which decays, from the leading-twist-2 term, as
  \begin{equation}
  \label{eq:largexbad}
    u^\pi(x)\sim (1-x)^1\;,
  \end{equation}  
  feeding the controversy provoked by the E615-Experiment leading order (LO) analysis~\cite{Conway:1989fs}, which favors a large-$x$ exponent of ``1'', in apparent contradiction with the parton models and pQCD. Many theoretical and phenomenological approaches have been considered in this debate, \emph{e.g.}~\cite{Holt:2010vj, Hecht:2000xa, Wijesooriya:2005ir, Aicher:2010cb,Chang:2014lva,Chen:2016sno, deTeramond:2018ecg,Cui:2020dlm,Ding:2019lwe,Ding:2019qlr, Sufian:2020vzb,Joo:2019bzr,Sufian:2019bol,Oehm:2018jvm,Brommel:2006zz,Detmold:2003tm}. Playing a key role in this controversy, the analysis of Aicher \emph{et al.}~\cite{Aicher:2010cb} shows that, if a next-to-leading order (NLO) treatment of the data is performed and soft-gluon resummation is considered, it is possible to recover the pQCD prediction. From a different perspective, the $x\to1$ profile of Eq.~\eqref{eq:largexgood} is also favored by a recent lattice QCD (lQCD) result~\cite{Sufian:2020vzb},  in which a novel ``Cross Section'' (CS) technique~\cite{Sufian:2019bol,Sufian:2020vzb} was employed to obtain the point-wise shape of the pion PDF.
  
  Furthermore, it is important to unravel the proton and pion properties together. Consider, for example, the origin and difference of their masses: if we accept Quantum Chromodynamics (QCD) as the fundamental underlying theory of the strong interactions (and we do), it is necessary to simultaneously explain the \emph{masslessness} of the pion and the much larger mass of the proton~\cite{Horn:2016rip,Roberts:2016vyn,Roberts:2019ngp,Roberts:2020udq}. Similarly, it is vital to obtain a clear picture of the proton and pion parton distributions in the same approach. QCD predicts the profiles of Eqs.~\eqref{eq:largexgood0}-\eqref{eq:largexgood}, thus we need to explain how those behaviors can (or cannot) take place. 
  
  In this manuscript, we revisit Ref.~\cite{deTeramond:2018ecg} to study the pion valence-quark PDF in the LFHQCD approach. Therein, the authors present an appealing way to parameterize the PDFs and generalized parton distributions (GPDs), starting from an integral representation of the EFFs. They claim that the falloff of the pion PDF at $x\to1$ is an unresolved issue. Our aim is to show that the large-$x$ behavior of Eq.~\eqref{eq:pionPDF1} can be perfectly accommodated within the same LFHQCD formalism, without compromising the EFF and also maintaining the correct counting rules for the proton.

 \noindent\emph{Counting rules in LFHQCD}\,---\, The hadronic form factor might be expressed in terms of an effective single particle density~\cite{Soper:1976jc}:
 \begin{equation}
 \label{eq:FF0}
 F(Q^2)=\int_0^1dy \;\rho(y,Q)\;,
 \end{equation}
 where $Q^2=-t$ is the photon momentum. The simplicity of the bulk-to-boundary propagators, in the soft-wall holographic model~\cite{Brodsky:2007hb}, enables us to obtain an analytically tractable expression for $\rho(y,Q)$~\cite{Brodsky:2007hb,Brodsky:2014yha,Zou:2018eam}. For arbitrary twist-$\tau$:
 \begin{equation}
\rho(y,Q)=(\tau-1)(1-y)^{\tau-2}y^{-t/4\lambda},
 \end{equation}
 such that, given the definition of Eq.~\eqref{eq:FF0}, the form factor can be simply expressed as:
  \begin{eqnarray}
 \label{eq:FF1}
 F_{\tau}(t)&=&(\tau-1)B(\tau-1,1-t/4\lambda)\;,
 \end{eqnarray}
where $B(u,v)$ corresponds to the Euler Beta Function (EBF), and $\lambda$ is a universal mass scale that will be defined later. Notice that if $\tau$ takes integer values (\emph{i.e.}, the anomalous dimensions are not taken into account), the EBF generates mass poles in the time-like axis. Those are eventually associated with the $\rho$-meson and its excitations, but at this point, the location of the poles is inadequate~\cite{Zou:2018eam}. A simple amendment consists in shifting the arguments of the EBF:
\begin{eqnarray}\nonumber
F_{\tau}(t)&=& \frac{1}{N_{\tau}}B\left(\tau-1,\frac{1}{2}-\frac{t}{4\lambda}\right)\\
\label{eq:EFFdef}
&=&\frac{1}{N_{\tau}}\int_{0}^{1} dy (1-y)^{\tau-2} y^{-t/4\lambda-\frac{1}{2}}\;,
\end{eqnarray}
where $N_\tau=\Gamma(1/2)\;\Gamma(\tau-1)/\Gamma(\tau-1/2)$. For integer values of $\tau$, it generates the following pole structure:
\begin{eqnarray}
\nonumber
F_\tau(t)\sim \frac{1}{(1-t/M_0^2)(1-t/M_1^2)\cdots (1-t/M_{\tau-2}^2)}\;,
\end{eqnarray}
with $M_n^2=4\lambda(n+1/2)$ and the universal scale $\lambda$ fixed by the $\rho$ meson mass~\cite{Brodsky:2014yha,Brodsky:2016yod}, $\sqrt{\lambda}=0.548$ GeV = $m_\rho/\sqrt{2}$. Thus, Eq.~\eqref{eq:EFFdef} corresponds to the integral representation of the form factor employed in Ref.~\cite{deTeramond:2018ecg}, which we exploit throughout this work. Under the change of variable $y=w_{\tau}(x)$ one can write, more generally: 
\begin{equation}
\label{eq:EFFdef2}
F_{\tau}(t)=\frac{1}{N_{\tau}}\int_{0}^{1} dx (1-w_{\tau}(x))^{\tau-2} w_{\tau}(x)^{-t/4\lambda-\frac{1}{2}}\frac{\partial w_{\tau}(x)}{\partial x}\, ,
\end{equation}
where the reparametrization function, $w_\tau(x)$, is constrained by the following conditions:
\begin{eqnarray}
\label{eq:const}
	w_{\tau}(0)=0;\,w_{\tau}(1)=1;\,\frac{\partial w_{\tau}(x)}{\partial x}\ge 0\, .
\end{eqnarray}
Notice that we have not ruled out a $\tau$-dependence in $w_\tau(x)$, which is a key difference with respect to~\cite{deTeramond:2018ecg}. The zero-skewness valence-quark GPD is conveniently expressed as
\begin{equation}
\label{eq:GPDF}
H_v^q(x,t):=H_v^q(x,\xi=0,t)=q_{\tau}(x)\text{e}^{t f_{\tau}(x)}\;,
\end{equation}
which follows from the definition of the flavor-$q$ form factor in terms of the GPD, $F^q(t)=\int_0^1H_v^q(x,t)$. From Eq.~\eqref{eq:GPDF} we identify the PDF and profile function, $q_\tau(x)$ and $f_{\tau}(x)$ respectively, thus
\begin{eqnarray}
\label{eq:pdfprof1}
q_{\tau}(x)&=&\frac{1}{N_{\tau}}(1-w_{\tau}(x))^{\tau-2} w_{\tau}(x)^{-\frac{1}{2}}\frac{\partial w_{\tau}(x)}{\partial x}\, ,\\
f_{\tau}(x)&=&\frac{1}{4\lambda}\text{log}\left(\frac{1}{w_{\tau}(x)}\right)\,.
\end{eqnarray}
Then, a simple form for $w_{\tau}(x)$ is suggested:
\begin{equation}
\label{eq:wexpl}
w_{\tau}(x)=x^{(1-x)^{g(\tau)}}\text{e}^{-a_\tau(1-x)^{g(\tau)}}\;,
\end{equation}
where $g(\tau),\;a_\tau\;\textgreater\; 0$.  It is straightforward to check whether the above expression meets the reparametrization invariance conditions of Eqs.~\eqref{eq:const}. First, an expansion of Eq.~\eqref{eq:wexpl} around $x=0$ yields
\begin{equation}
w_{\tau}(x)=x\;[1+g(\tau)(a_\tau - \ln (x))x+\mathcal{O}(x^2) ]e^{-a_\tau}\;,
\end{equation}
which implies $w_{\tau}(0)=0$. Taking the logarithm in Eq.~\eqref{eq:wexpl}, one obtains
\begin{eqnarray}
\ln [w_{\tau}(x)]&=& (1-x)^{g(\tau)}[\ln(x)-a_\tau] \;,
\end{eqnarray}
such that $\ln [w_{\tau}(1)] = 0$, from which $w_{\tau}(1)=1$. Finally, the derivative constraint in Eqs.~\eqref{eq:const} can be checked by applying the chain rule on
\begin{eqnarray}
\frac{\partial \ln [w_{\tau}(x)]}{\partial x}=\frac{1}{w_\tau(x)}\frac{\partial w_{\tau}(x)}{\partial x}\;,
\end{eqnarray}
 noticing that $w(x)\textgreater 0$ and $\frac{\partial \ln [w_{\tau}(x)]}{\partial x} \ge 0$, when $x \in [0,1]$. 

At this point it, is worth stressing that $w_{\tau}(x)$ is neither unique nor derived from first principles; however, its particular form can be motivated by both mathematical and physical constraints. Our proposed profile for $w_\tau(x)$ adopts the desired Regge behavior at small-$x$~\cite{Zou:2018eam,deTeramond:2018ecg}, while also satisfying the constraints of Eqs.~\eqref{eq:const}, which are necessary for the reparametrization invariance of the EBF. This invariance property ensures that the form factors we obtain are identical to those from Ref.~\cite{deTeramond:2018ecg}, thus preserving the good properties, including the wanted large-$t$ falloff:
\begin{equation}
\label{eq:EFFlarget}
F_{\tau}(Q^{2})\sim \left(\frac{1}{-t}\right)^{\tau-1}\;,
\end{equation}
that is, the correct power-law asymptotic behavior of the form factor~\cite{Lepage:1980fj,Farrar:1975yb,Ezawa:1974wm} is faithfully reproduced. Focusing on the valence-quark PDF, Eqs.~\eqref{eq:pdfprof1}-\eqref{eq:wexpl} imply that the $x\to1$ leading power of $q_{\tau}(x)$ exhibits the following $\tau$-dependence:
\begin{equation}\label{eq:PDFdef}
q_{\tau}(x)\sim (1-x)^{h(\tau)}\;,
\end{equation}
with $h(\tau)=(\tau-1)g(\tau)-1$. Due to the arbitrariness of the choice of $g(\tau)$, LFHQCD cannot predict its precise form or, consequently, the exact counting rules. However, it is this flexibility that allows us to recover the corresponding counting rules for both pion and proton, Eqs.~\eqref{eq:largexgood0}-\eqref{eq:largexgood}. Given the simplicity of Eq.~\eqref{eq:PDFdef}, we propose the following rules for the PDFs:
\begin{eqnarray}
\label{eq:Rules}
 \text{Rule-I}&:& (1-x)^{2\tau-3} \,,\; \text{with} \, g(\tau)=2\;. \\
 \text{Rule-II}&:& (1-x)^{2\tau-2} \,,\; \text{with} \, g(\tau)=2+\frac{1}{\tau-1} \;.
\end{eqnarray}
Thus, it is inferred from Eq.~\eqref{eq:PDFdef} that the spin$-\frac{1}{2}$ relation, Eq.~\eqref{eq:PDFDYW}, can be satisfied if Rule-I is chosen, while the spin$-0$ counterpart, Eq.~\eqref{eq:pionPDF1}, holds if Rule-II is selected instead. The effects of applying these rules on the pion valence-quark PDF will be tested numerically in the following section.

  \begin{figure}[t]
	\centering
	\includegraphics[width=0.95\columnwidth]{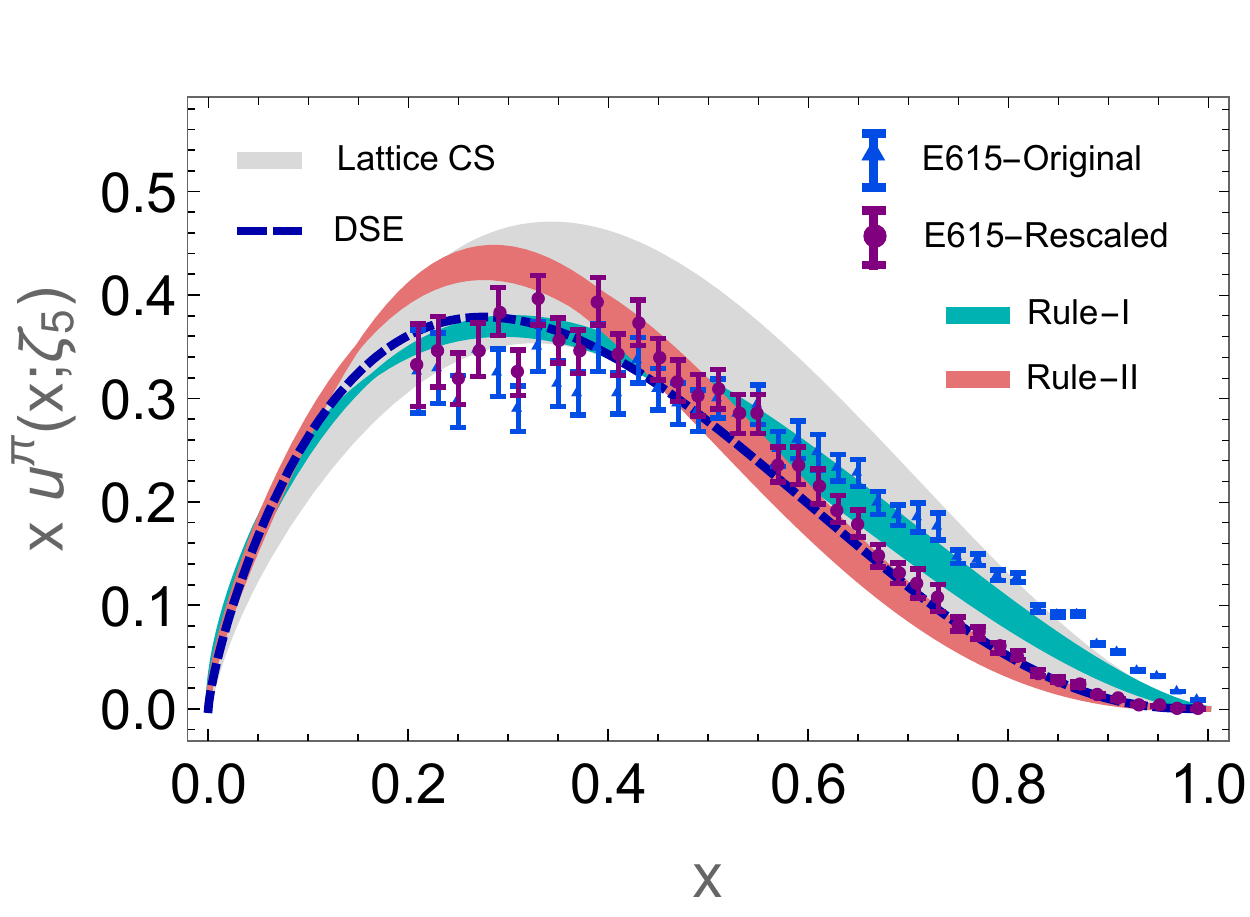}
	\caption{\textit{Valence-quark pion PDF.} Obtained NLO results at $\zeta_5=5.2$ GeV, from  the rules in Eqs.~\eqref{eq:Rules}. The corresponding (blue and red) error bands account for the uncertainty in the initial scale, $\zeta_1=1.1\pm0.2$ GeV and the variation of $a_4/a_2 = 0.1$ to $1$. The broadest, gray band, corresponds to the novel lQCD ``CS'' result from~\cite{Sufian:2020vzb} and the dashed-line depicts the DSE prediction~\cite{Ding:2019lwe,Ding:2019qlr}. \textbf{Data points:} (triangles) LO extraction ``E615-Original''~\cite{Conway:1989fs} and (circles) the NLO analysis ``E615-Rescaled'' of Ref.~\cite{Aicher:2010cb}.    }
	\label{fig:PDF}
\end{figure}

\begin{figure}[t]
	\centering
	\includegraphics[width=0.95\columnwidth]{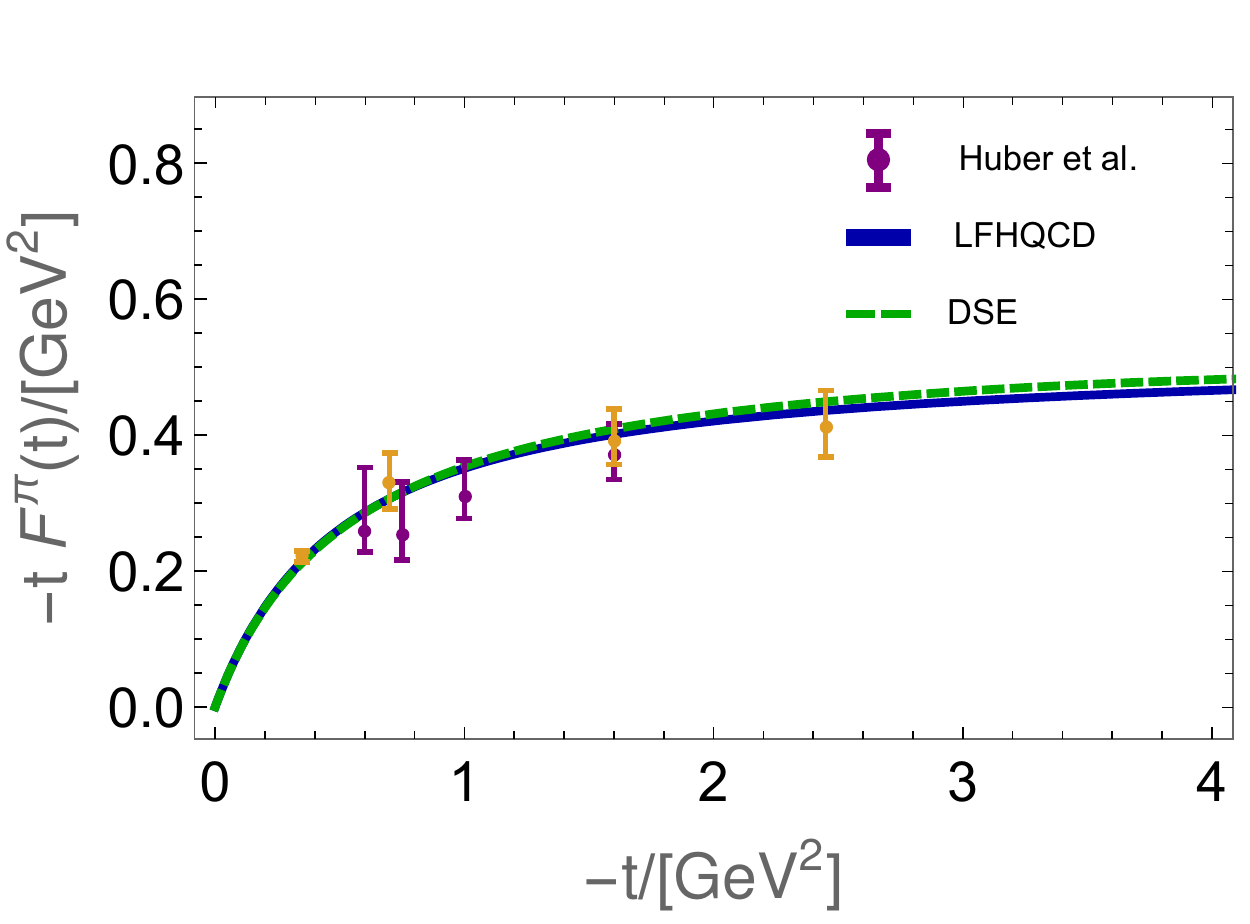}
	\caption{\textit{Pion form factor.} LFHQCD result and its comparison with the DSE prediction~\cite{Chang:2013nia} and experimental data~\cite{Huber:2008id}. Interestingly, the LFHQCD and DSE results lie almost on top of each other. }
	\label{fig:EFF}
\end{figure}

\begin{figure}[t]
	\centering
	\includegraphics[width=0.95\columnwidth]{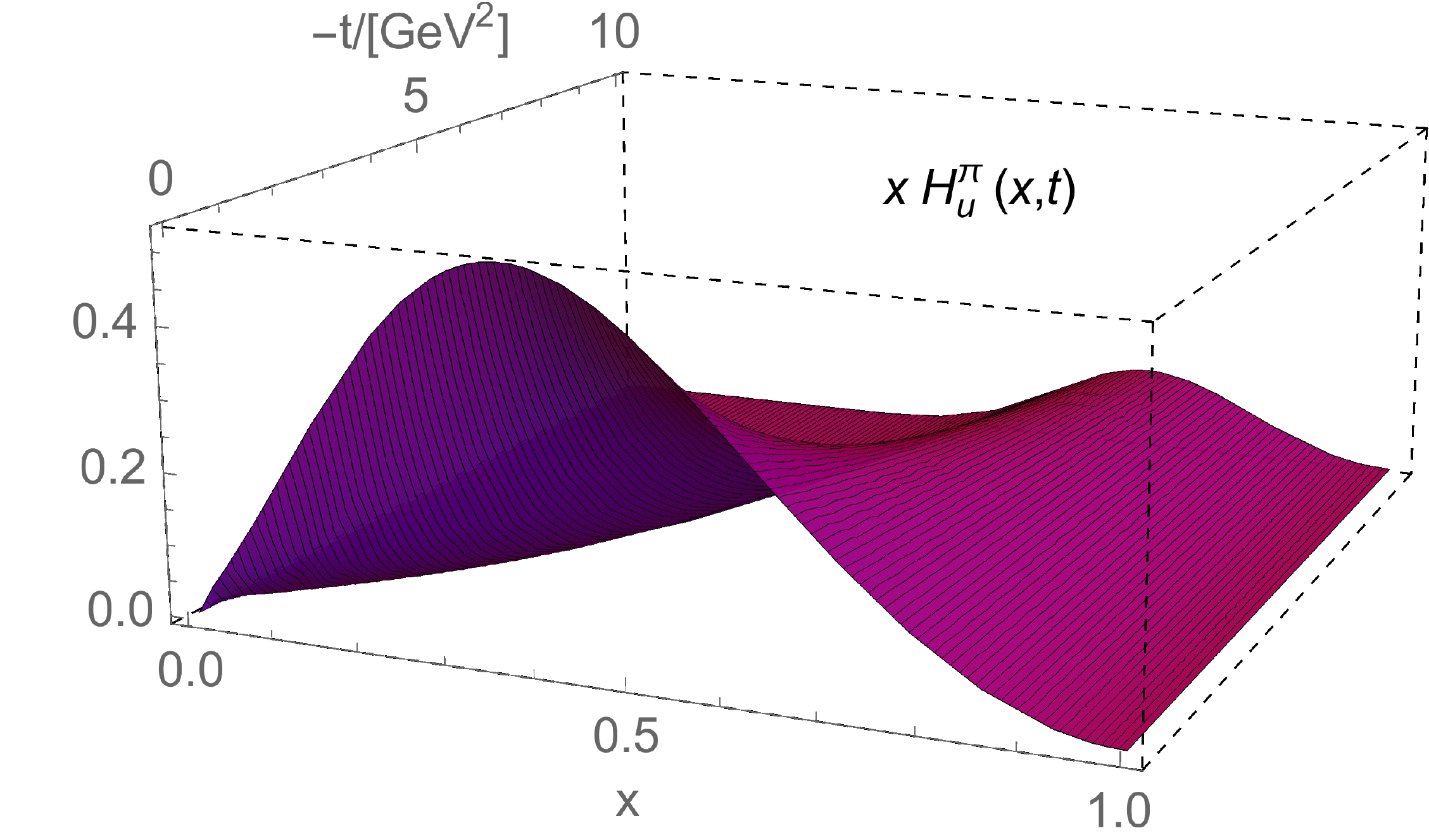}
	\caption{\textit{Valence-quark pion GPD.} $t$-dependence of the zero-skewness valence-quark GPD, $H_u^\pi(x,t)$. The plot above corresponds to Rule-II in Eq.~\eqref{eq:Rules}, at the initial scale $\zeta_1$.   }
	\label{fig:GPD}
\end{figure}

 \noindent\emph{Pion valence-quark PDF}\,---\,
Consider the twist-4 pion valence-quark PDF as
\begin{equation}
\label{eq:PDFtwist4}
u^{\pi}(x;\zeta)=(1-\gamma)q_{\tau=2}(x;\zeta)+\gamma q_{\tau=4}(x;\zeta)\;,
\end{equation}
with normalization $\int_0^1 dx \;u_\pi(x;\zeta)=1$ and $\gamma=0.125$. The parameter $\gamma$ controls the strength of the twist-$4$ component. It is fixed by the meson cloud contribution determined in~\cite{Brodsky:2014yha}. The PDF is defined at an intrinsic scale $\zeta=\zeta_1$, which is set as $\zeta_1=1.1\pm 0.2$ GeV to keep in line with previous works~\cite{deTeramond:2018ecg,Deur:2016opc}. Then, continuum analyses~\cite{Cui:2020dlm,Ding:2019lwe,Ding:2019qlr} are employed for benchmarking, to estimate 
\begin{eqnarray}
\textless x \textgreater_{\zeta_1}^u=\int_0^1 dx\;x u^\pi(x;\zeta_1)\approx 0.26\;,
\end{eqnarray}
 such that the $a_2$ coefficient in Eq.~\eqref{eq:wexpl} can be determined. This is additionally cross-checked from the value $\textless x \textgreater_{\zeta_2}\approx 0.24$, obtained at $\zeta_2:=2$ GeV after NLO evolution, as compared to the lQCD estimates from Refs.~\cite{Joo:2019bzr,Detmold:2003tm,Oehm:2018jvm,Brommel:2006zz}. Furthermore, we find this result compatible with a recent determination from the xFitter collaboration~\cite{Novikov:2020snp}. To account for the impact of the twist-4 term, we also vary the ratio $a_4/a_2$ from $0.1$ to $1$. Only mild effects at intermediate values of $x$ are observed. Figure~\ref{fig:PDF} displays the resulting valence-quark PDFs, evolved to $\zeta_5:=5.2$ GeV, and its comparison with experimental and lattice data~\cite{Conway:1989fs,Aicher:2010cb,Sufian:2020vzb}. For contrast, we have also included a recent results from Dyson-Schwinger equations (DSEs) framework~\cite{Ding:2019lwe,Ding:2019qlr}. The pion form factor is shown in Figure~\ref{fig:EFF}; we compare with the JLab experimental results~\cite{Huber:2008id} and with the DSE prediction from Ref.~\cite{Chang:2013nia}. In addition, the corresponding zero-skewness valence-quark GPD, for Rule-II, is depicted in Figure~\ref{fig:GPD}.
 
It is clear that Rule-I produces a PDF that closely corresponds to the \emph{original} analysis of the experimental data~\cite{Conway:1989fs}, while the analogous for Rule-II  matches the \emph{rescaled} data from Ref.~\cite{Aicher:2010cb}. Both rules produce the same well-behaved EFF, with the correct large-$t$ power-law decay,  but only  in the second case one obtains  the $x\to 1$ behavior of the PDF predicted by pQCD. This is readily achieved in the DSE formalism~\cite{Cui:2020dlm,Ding:2019lwe,Ding:2019qlr}: its direct connection with QCD ensures that the perturbative limits are recovered, and so the relation between the asymptotic behavior of the gluon with the large-$x$ profile of the valence-quark PDF~\cite{Holt:2010vj}. Moreover, state-of-the-art lQCD results~\cite{Sufian:2020vzb} on the point-wise form of the PDF also establish that the asymptotic form of Eq.~\eqref{eq:largexgood} is preferred.

\begin{figure}[t]
	\centering
	\includegraphics[width=0.95\columnwidth]{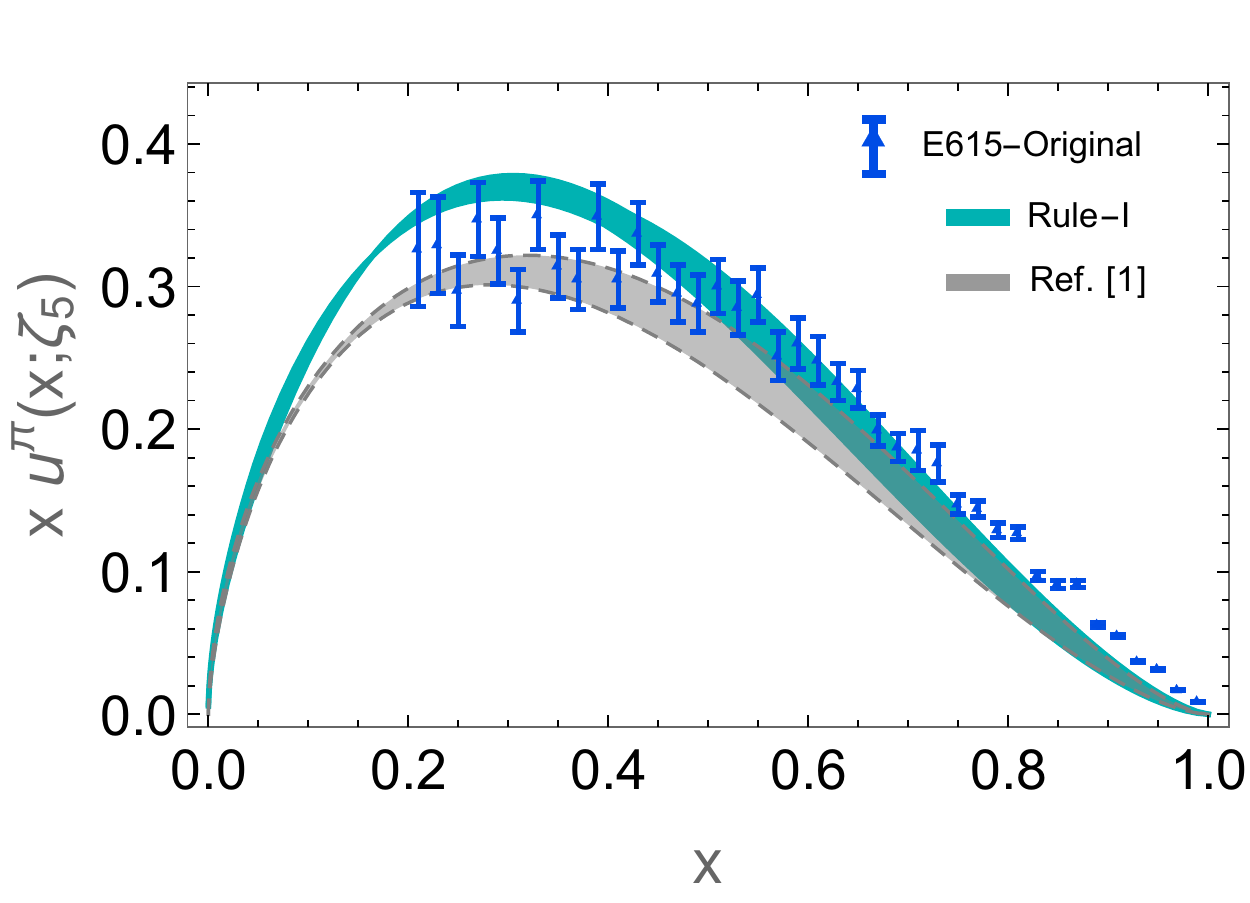}
	\caption{\textit{Valence-quark pion PDF.} Obtained NLO results at $\zeta_5=5.2$ GeV, for  Rule-I in Eqs.~\eqref{eq:Rules}, and its comparison with the prediction of Ref.~\cite{deTeramond:2018ecg}. In both cases, the large-$x$ exponent of ``1'' prevails. The corresponding error bands account for the uncertainty in the initial scale, $\zeta_1=1.1\pm0.2$ GeV; our result also considers the variation of $a_4/a_2 = 0.1$ to $1$. \textbf{Data points:} (triangles) LO extraction ``E615-Original''~\cite{Conway:1989fs}. }
	\label{fig:PDF1}
\end{figure}

 It is noteworthy that, even though the pion valence-quark PDF obtained from Rule-I differs from that computed in~\cite{deTeramond:2018ecg}, the evolved results are compatible, as can be seen in Figure~\ref{fig:PDF1}. This is unsurprising since the corresponding reparametrization function is not dramatically different from its counterpart in~\cite{deTeramond:2018ecg}. The most noticeable differences occur at an intermediate range of $ x $, around $ x \approx 0.1 - 0.4 $, but the PDFs are quite similar outside this domain (thus the large-$x$ behavior is not compromised). We would expect something similar for the case of the proton. Moreover, as explained in the previous section, owing to the reparametrization invariance of the EBF, our Rule-I and the counting rule from~\cite{deTeramond:2018ecg} produce the same proton EFF. Thus, although it is not included in the present manuscript, we expect Rule-I to produce a realistic picture for the proton. These observations encourage us to select Rule-I for the case of the proton and Rule-II when studying pions, for an internally consistent description based on the LFHQCD formalism. 

 \noindent\emph{Summary and conclusions}\,---\,
 We have reanalyzed the LFHQCD approach of Ref.~\cite{deTeramond:2018ecg} for the pion valence-quark PDF, $u^\pi(x)$. It has been proven that given the flexibility of the reparametrization function, $w_\tau(x)$, it is in fact possible to accommodate a large-$x$ behavior of $u^\pi(x)\sim(1-x)^{2\tau-2}$ within this framework.  In addition to the agreement with the rescaled experimental data~\cite{Aicher:2010cb}, our conclusions on $u^\pi(x)$ are compatible with the Ezawa findings~\cite{Ezawa:1974wm} and the predictions from pQCD~\cite{Farrar:1975yb,Berger:1979du,Lepage:1980fj}.  Recent continuum~\cite{Ding:2019lwe,Ding:2019qlr} and sophisticated lQCD studies~\cite{Sufian:2020vzb} also favor this endpoint form. Owing to this confluence of vastly different approaches, and given our observations, we state that the $u^\pi(x)\sim(1-x)^2$ profile can not only be contained within the LFHQCD formalism, but also cannot be excluded. This has also been explored recently in the related approach of AdS/QCD~\cite{Lyubovitskij:2020otz}. Notably, the form factors remain unaltered regardless of the chosen rule, as a consequence of the reparametrization invariance of the EBF. Thus, the pion EFF exhibits a remarkable agreement with the experimental data and DSE predictions, while also manifesting the correct power law at large-$t$. Our Rule-I result for the pion valence-quark PDF, at the experimental scale, differs moderately from the one obtained in~\cite{deTeramond:2018ecg} in a limited domain, $x\approx 0.1 - 0.4$; outside this range, in particular the large-$x$ regime, the agreement is perfect. Therefore we expect something similar to happen for the proton case. With these ideas in mind, we can sketch how a concurrent description of the proton and pion distribution functions, which agrees with pQCD, can be achieved within this formalism if the counting rules are chosen accordingly: we encourage the use of Rule-I for the proton and Rule-II for the pion.

\medskip
We acknowledge helpful conversations with Yuan Sun.
This work is supported by: the Chinese Government Thousand Talents Plan for Young Professionals.

\bibliographystyle{unsrt}
\bibliography{bibliography}

\end{document}